\documentclass[final,english]{bullsrsl}[2022/06/15]

\usepackage[latin1]{inputenc}
\usepackage[T1]{fontenc}
\vspace{50px}
\usepackage{natbib} 
\usepackage{graphicx}
\begin{document}
\title{UVIT/AstroSat Investigation of a low luminous Blue Straggler Star in NGC 362: Detection of extremely low mass white dwarf as companion }

\author[affil={1,2}, corresponding]{Arvind K. }{Dattatrey}
\author[affil={2}, corresponding]{R. K. S.}{Yadav}
\author[affil={3}]{Annapurni}{Subramaniam}
\author[affil={2}]{Ravi S.}{Singh}
\affiliation[1]{Aryabhatta Research Institute of Observational Sciences, Manora Peak Nainital, 263001, India.}
\affiliation[2]{ Deen Dayal Upadhyay Gorakhpur University, Gorakhpur, Uttar Pradesh 273009, India.}
\affiliation[3]{Indian Institute of Astrophysics, Koramangala, Bangalore 560034, India.}

\correspondance{arvind@aries.res.in,rkant@aries.res.in}
\date{31st MAy 2023}
\maketitle


%
\begin{abstract}

In the present study, we identified an extremely low-mass white dwarf  as a companion to a low luminous  blue straggler star  within the Galactic globular cluster NGC 362. To conduct the analysis, we utilized data obtained from various sources, including AstroSat's Ultra Violet Imaging Telescope, UVOT, and the 2.2-m ESO telescope. By examining the spectral energy distribution  of the  blue straggler star candidate, we successfully identified an extremely low-mass white dwarf as its binary companion.  We determined the effective temperature, radius, and luminosity for both, the extremely low-mass white dwarf and the blue straggler star candidate. Specifically, the extremely low-mass white dwarf  exhibits an effective temperature of 15000 K, a radius of 0.17 R${\odot}$, a luminosity of 1.40 L${\odot}$, and a mass range of 0.19-0.20 M$_{\odot}$.  Furthermore, the position of the straggler star within the cluster suggests their formation via the Case A/B mass-transfer mechanism in a low-density environment.
    
\end{abstract}


\section{INTRODUCTION}
Globular clusters (GCs) were among the earliest known celestial systems in which stars regularly interacted gravitationally. These interactions result in diverse phenomena, including two-body relaxation, mass segregation from core collapse, energy equipartition, stellar collisions, mass transfer, and binary star mergers. Gravitationally interacting binary stars can endure this process, in which the more massive star consumes its companion, growing in size and luminosity until it transforms into a blue straggler star (BSS). 

Extensive research has been conducted on white dwarfs (WDs), which possess helium cores and are commonly referred to as extremely low-mass white dwarfs (ELM WDs). These unique WDs have masses up to approximately 0.20 M$_{\odot}$. It is believed that white dwarfs resulting from the evolution of isolated stars have a maximum mass of around 0.4  M$_{\odot}$ which is connected to the age of the universe \citep{2010ApJ...723.1072B}. Compact binary systems have been observed to contain white dwarfs with masses ranging from 0.1 to 0.4  M$_{\odot}$ \citep{2004MNRAS.352..249B,2010ApJ...723.1072B}. These white dwarfs are thought to have formed due to mass loss in these binary systems.
   
The GC NGC 362 is located at the celestial coordinates (RA = $01^h\, 03^m\, 14.26^s$ and Dec = $-70^{\circ}$  50$'$ 55$''$.6) in the constellation Tucana, situated in the Southern hemisphere. Estimates suggest that this cluster has an age of approximately 11  Gyr \citep{2021MNRAS.505.5978V} and is located at a distance of  8.83 kpc from Earth. It exhibits a reddening value of $~$$\sim$ 0.05 mag and a metallicity [Fe/H] of  $\sim$  -1.3 dex \citep{2010arXiv1012.3224H}. 

\section{DATA SETS AND THEIR REDUCTION}
\label{sec:data}
The GC NGC 362 is analyzed using images captured by the UVIT ( \cite{2017JApA...38...28T}  ) instrument on the AstroSat satellite. The observations took place on November 11, 2016, using four UV bands: F148W, F169M, N245M, and N263M.  In addition to the UVIT data, we used Ultraviolet Optical Telescope (UVOT) data. The raw data from the UVOT, sourced from the HEASARC archive, were processed using the HEA-Soft pipeline \citep{2008MNRAS.383..627P}. The detailed information about the UVOT instrument and the photometric correction can be found in  \cite{2008MNRAS.383..627P}. Furthermore, we included optical data from past observations conducted with the Wide-Field Imager (WFI) in U, B, V, R, and I bands mounted on the 2.2 m ESO/MPI Telescope. Detailed information about data and photometry has also been given in \cite{2023ApJ...943..130D}.
\section{UV AND OPTICAL COLOR-MAGNITUDE DIAGRAMS}
\label{sec:cmds}
We conducted  the present analysis by cross-matching the UVIT data with the Gaia EDR3 photometric data within a matching radius of $1''.5$. The membership probability of the stars  is more than 80\%, taken from \cite{2022MNRAS.514.1122S}. The membership probability has been determined using GAIA DR3 proper motions data. It is based on the comparison of the cluster-like stars' kinematics distribution and the field distribution. We created the UV-optical color-magnitude diagrams (CMDs) by considering the optical counterparts of UVIT detected sources.

\begin{figure}    
\hspace*{-0.15cm}
    \includegraphics[scale=0.225]{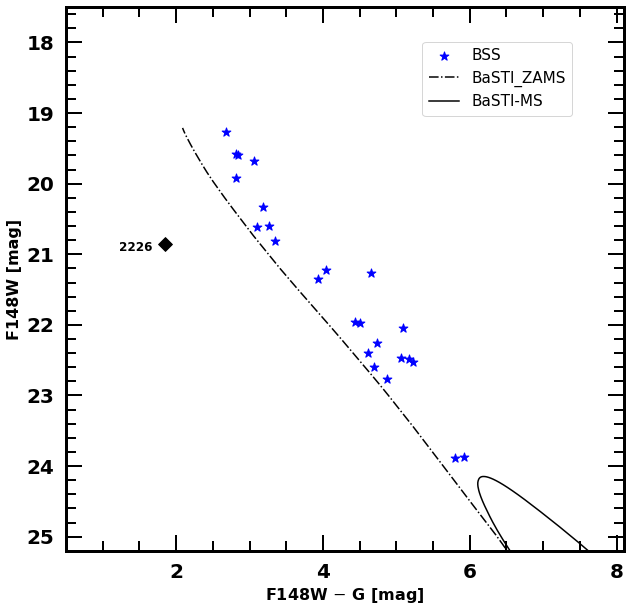}
    \includegraphics[scale=0.22]{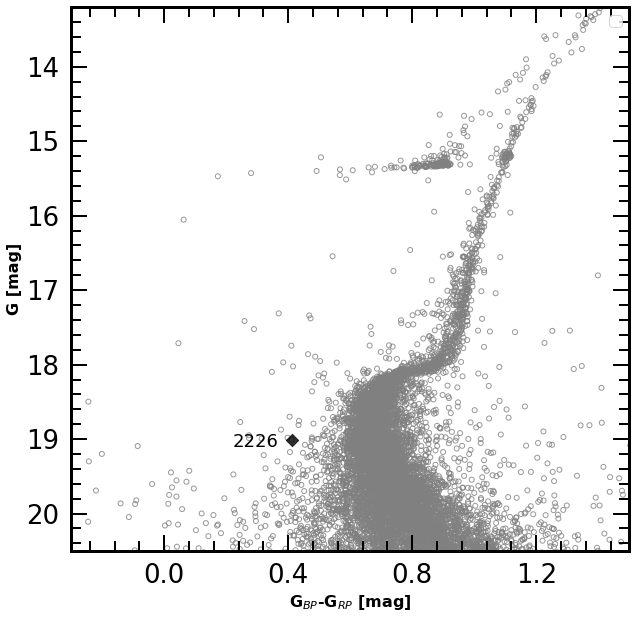}
    \includegraphics[scale=0.22]{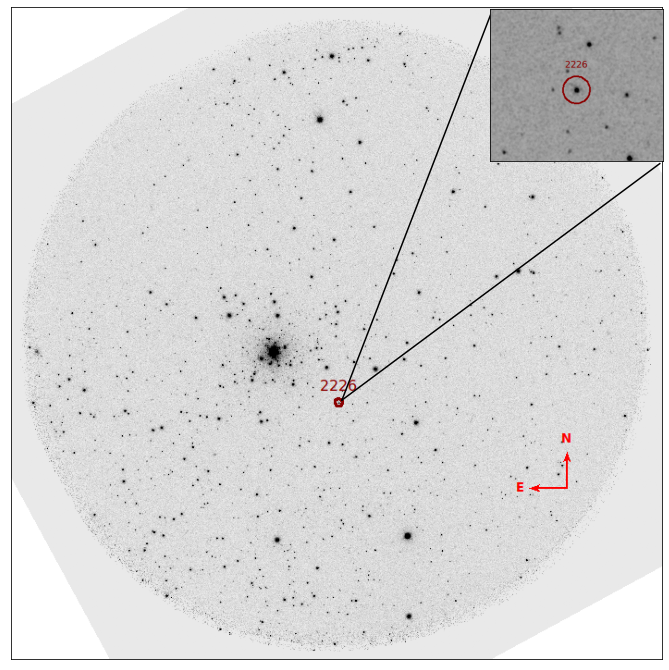}
    \caption{Left panel: UV-optical CMD (F148W vs F148W $-$ G) of the BSSs for the cluster NGC\,362. Blue symbols represent the BSSs. The probable Blue straggler candidate, bluer than the ZAMS, is displayed with a  black diamond symbol.\label{fig:FUV_G_CMD}  Middle panel: The optical CMD (G vs G$_{BP}$ $-$ G$_{RP}$) of the cluster NGC\,362.  Right panel: Spatial locations of an FUV bright BSS candidate with a red circle in the F148W image of NGC\,362. The field of view of the image is $\sim$28$'$ x 28$'$. The zoomed region around the target star is shown in the top right corner.} North is up and east is on the left.
    \label{fig:CMD_gaia}
\end{figure}

To select the BSSs, we adopted the method given in \cite{2023ApJ...943..130D}. Basically, we made a box around the BSS sequence using  CMD. To determine the bottom limit of the box, we computed the average magnitudes and colors with standard deviation ($\sigma$) of stars within one magnitude box near the main sequence (MS) turn-off point. The lower limit of the box is greater than  5$\sigma$  from the MS turn-off. Our primary focus was on the FUV optical CMD (F148W, (F148W $-$ G)), which provided crucial information about the cluster stars. In Figure \ref{fig:CMD_gaia} (left panel), we plotted the CMD, where blue  symbols   represent the BSSs.  To  interpret the CMD, we introduced dotted and solid lines representing the Zero-Age Main Sequence (ZAMS) and Main Sequence  curves derived from BaSTI\footnote{http://basti-iac.oa-abruzzo.inaf.it/}.  We have found an FUV bright star below the ZAMS indicated by  no. 2226 and shown with a black diamond. To cross-check the position of the star 2226 in the optical CMD, we created a (G, (G$_{BP}$ $-$ G$_{RP}$)) diagram, as shown in the middle panel of Figure \ref{fig:CMD_gaia}.  The diagram clearly indicates that star 2226 aligns below the BSS sequence of the cluster.
In order to validate the position of the star 2226 and assess any potential contamination, we carefully inspected the observed image. The right panel of Figure \ref{fig:CMD_gaia}  depicts the F148W image, where we marked the position of the star 2226 with a red square. The image reveals that this star is well-resolved and exhibits no signs of contamination. 

As BSSs are often formed through interactions between stars, it is plausible that some BSSs have a companion in the form of a binary system. This companion could exist either as the outer component of a triple system or as a remnant of a donor star after Mass Transfer (MT) has occurred. Analyzing the SEDs of these systems allows us to determine the characteristics of their multiple components. The successfully detected companions known as ELM WDs associated with BSSs indicate that these systems have undergone MT \citep{2023ApJ...943..130D}. 

\section{SPECTRAL ENERGY DISTRIBUTION}
SED modeling involves fitting theoretical models or templates to the observed data to derive the physical parameters of the object, such as its temperature, luminosity, and radius. As a virtual observatory tool, VOSA (VO SED Analyzer) \citep{2008A&A...492..277B} is used in this work.  The produced synthetic photometric data is compared with the observed data to find the best fit using the $\chi^2$ minimization test.  We used the \cite{2003IAUS..210P.A20C}  model with a full range of all parameters such as the effective temperatures (T$_{eff}$) from 5000-50000 K, with a metallicity of [Fe/H] = -1, which is close to the metallicity of the cluster, and a log $g$ of 3-5 dex provided in VOSA for the cool component. To account for the UV excess observed in the data, we employed two-component SED.  For the hotter component, we used the Koester model \citep{2010MmSAI..81..921K} with a full range of parameters such as the T$_{eff}$ and log $g$ range from 5000$-$80000 K and 6.5$-$9.5, respectively.

\begin{figure}
\includegraphics[scale=0.23]{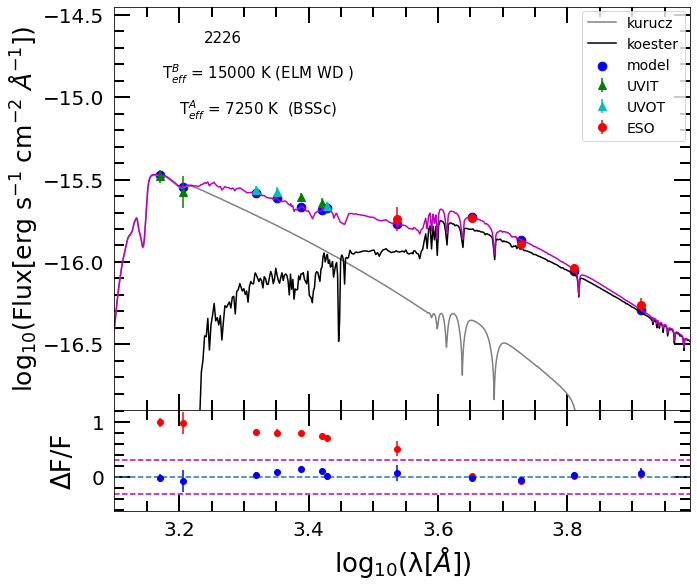}
\includegraphics[scale=0.34]{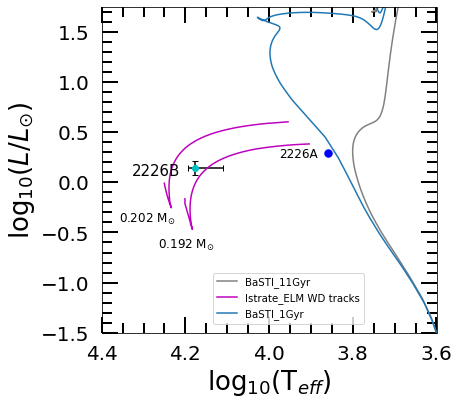}
\caption{Left panel, the SED of  bright BSS candidate star. The T$_{eff}$ of the cool (A) and hot (B) components are displayed in SED. The observed data points are represented by green, cyan, and red  points representing UVIT, UVOT, and ESO data respectively, while blue points represent the model points. Black,  grey, and magenta curves represent the Koester, Kurucz, and combined spectra, respectively. $\Delta F/F$ is the fractional residual. Right Panel, the HR diagram of the cool (blue) and hot (cyan) components. The grey curve represents the BaSTI isochrones. The ELM tracks are represented by the magenta curves with masses around 0.192 and 0.202M$_{\odot}$.}
\label{fig:SED}
\end{figure}

First, we fitted a SED with a single component, as depicted by the grey curve in the left panel of Figure \ref{fig:SED}. The data points used for fitting are depicted as green, cyan, and red-filled circles with error bars, whereas model data points are depicted as blue points. The residual between the fitted model and the observed fluxes, normalized by the observed flux, is displayed in the bottom panel of Figure \ref{fig:SED} (left panel). The dashed horizontal lines at 0.3 (30\%) represent the threshold. The residual plot for the star 2226 reveals that the UV residual flux is more significant than 0.3 for multiple data points. This indicates that there is an UV excess, and the SED may be fitted by a combination of hot and cool spectra. Therefore, a two-component SED  is fitted, as depicted in Figure \ref{fig:SED} (left panel). The combined spectrum  is shown with a magenta line.  The fitting parameters for combine spectra such as  $\chi^2$, Vgf,  and Vgf$_{b}$  are 22.63, 19.0, and 4.0 respectively.

The fractional residual points for the combined SED is well fitted with 0.3 shown by blue points. The Fundamental properties of the star 2226, including the T$_{eff}$, log g, luminosity, and radius of the cool component (2226A) are estimated to be 7250$^{+125}_{-125}$ K, 5, 1.97$^{+0.181}_{-0.020}$ L$_{\odot}$, and 0.89$^{+0.03}_{-0.06}$ R$_{\odot}$, whereas for the hot component (2226B) are 15000$^{+250}_{-1000}$ K, 9.5, 1.40$^{+0.09}_{-0.10}$ L$_{\odot}$, and 0.17$^{+0.03}_{-0.01}$ R$_{\odot}$.

\subsection{HERTZSPRUNG-RUSSELL DIAGRAM}
To estimate the age and mass of the cool component (2226A), we utilized the Hertzsprung-Russell (H-R) diagram shown in Figure \ref{fig:SED}. By comparing the position of the cool component, represented by the blue point, with theoretical isochrones retrieved from BaSTI \citep{2021ApJ...908..102P}, we can determine their age and mass. In the H-R diagram, we plotted the 1 Gyr isochrone using the blue line. By evaluating the position of the cool component (2226A), we estimated the age and mass of BSSc as  $\sim 1$ Gyr  and 1.77 M$_{\odot}$, respectively.

For the hot components (2226B), we also determined the age and mass  using the H-R diagram displayed in the right panel of Figure \ref{fig:SED}. By considering the surface temperature and luminosity derived from the SED analysis, we positioned the hot component (2226B) with cyan point. By examining the position of the  hot component relative to the ELM tracks, we estimated cooling age and mass to be $\sim 4$ Myr and 0.19-0.20 M$_{\odot}$, respectively. 
  
  \section{Results and discussion}\label{sec:results}
In the present study, we  identified an FUV bright BSSc in the globular cluster NGC 362. We constructed SED using the VOSA tool to analyze the physical characteristics of  BSSc and investigate the presence of any companion. Fitting single-component SEDs to the observed data points revealed UV excess in the BSSc, indicating the need for two-component SED fitting. By fitting the SED with combined spectra of cool and hot components, we achieved good fits for BSSc. This suggest that  BSSc is part of the binary system. The hot component denoted as 2226B, exhibited a surface temperature  of 15000 K, luminosity of 1.40 L$_{\odot}$, and radii of 0.17 R$_{\odot}$, while for cool component,  it is 7250 K, 1.97 L$_{\odot}$, and 0.89 R$_{\odot}$ respectively. Based on the above parameters, we suggests that cool component may be BSSc  and hot component can be ELM WD. 

Based on the literature survey, the ELM WD can be found in binaries \citep{2023ApJ...943..130D}, indicating the formation process via  mass transfer (MT) processes. The presence of low-mass WDs in tight binary systems suggests that the companion star removes the progenitor's envelope, leading to the formation of low-mass WDs without igniting the helium core. In this case of BSSc, this system likely formed through MT with a ELM WD companion. The formation of this binary system may be related to the dynamical evolution of the cluster. The cooling age of the ELM WD indicates that this system  has recently formed. Binary systems such as BSS+ELM WDs are uncommon in GCs. \citet{2023ApJ...943..130D} identified 12 such systems in the GC NGC 362 for the first time. UV data are highly valuable in identifying such systems.

\section{Summary and Conclusions}
\label{sec:concludions}
The UVIT, UVOT, Gaia EDR3, and the 2.2m ESO/MPI telescopes provide data of  BSSc in NGC 362. We found a BSSc  with an ELM WD  companion. Based on their location in the HR diagram, the hot and cool components are  classified as ELM WD and BSS,c respectively.  The ELM WD accompanied by BSSc formed via the Case A/B mass transfer pathway. 

The telescopes under BINA collaborations are extremely helpful for the type of analysis presented here. The 3.6m optical measurements are critical for the SED study. The spectroscopic observations are quit useful for characterising the BSS in the GC's outer area.

\begin{furtherinformation}
\begin{authorcontributions}
This work is part of a long-term survey program and collective efforts were made by all the co-authors with the relevant contributions.
\end{authorcontributions}
\section{ACKNOWLEDGEMENTS}
I express my gratitude to the reviewer for their valuable and thoughtful comments and suggestions, which significantly enhanced the quality of this manuscript.

This publication extensively utilizes data from the Astrosat mission, conducted by the Indian Space Research Organisation (ISRO) and archived at the Indian Space Science Data Centre (ISSDC). The research also leverages VOSA, a tool developed as part of the Spanish Virtual Observatory project and supported by the Spanish Ministry of Economy and Competitiveness (MINECO) through grant AyA2017-84089.

Moreover, this study relies on the software tools TOPCAT (Taylor 2005, 2011), matplotlib (Hunter 2007), and NUMPY (van der Walt, Colbert Varoquaux 2011) for data analysis and visualization. Additionally, the NASA Astrophysics Data System (ADS NASA) plays a crucial role in supporting the research.
\begin{conflictsofinterest}
The authors declare no conflict of interest.
\end{conflictsofinterest}

\end{furtherinformation}
\bibliographystyle{bullsrsl-en}
\bibliography{main}

\end{document}